\documentclass[conference]{IEEEtran}

\usepackage[justification=centering]{caption}

 \usepackage[pdftex]{graphicx}

\begin{document}
%
\title{The Eve of 3D Printing in Telemedicine: State of the Art and Future Challenges}

\author{\IEEEauthorblockN{Piero Giacomelli}
\IEEEauthorblockA{IT Department\\
Spac S.p.A.\\
36071 Arzignano Italy\\
Email: giacomellip@spac-spa.it}
\and
\IEEEauthorblockN{\AA sa Smedberg}
\IEEEauthorblockA{The Department of Computer and Systems Sciences\\
Stockholm University \\
Kista, Sweden\\
Email: asasmed@dsv.su.se}
}


%


\maketitle

\begin{abstract}
3D printing has raised a lot of attention from fields outside the manufacturing one in the last years. In this paper, we will illustrate some recent advances of 3D printing technology, applied to the field of telemedicine and remote patient care. The potentiality of this technology will be detailed without lab examples. Some crucial aspect such as the regulation of these devices and the need of some standards will also be discussed. The purpose of this paper is to present some of the most promising applications of such technology.
\end{abstract}


\IEEEpeerreviewmaketitle

\section{Introduction}
3D printing technology is changing manufacturing models so fast that traditional industrial processes are chasing this new wave in a way very similar to the paradox described by the "Red Queen effect" \cite{ECOJ:ECOJ1002}. The first 3D printer was designed as early as in 1984 by Charles W. Hull (see Fig. 1). 

\begin{figure}[htbp]
	\centering
		\includegraphics[width=7.5cm]{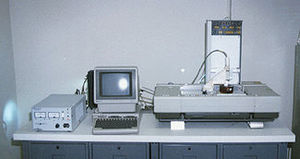}
		\caption{Early version of a 3D printer}
		\label{figure:1}
\end{figure}

However, it is only in the last ten years that the use of 3D printing technology 
outside the traditional manufacturing environment has started to 
revolutionize the way we traditionally turn raw materials into 
functional devices. The growing wave of 3D printing technology 
has been possible because of two important factors:

\begin{itemize}
\item{3D printing technology raised the critical mass that 
was needed to make manufacturers willing to sell their 
product to individuals and not solely to private companies. }
\item{the technology started to become available on the 
World Wide Web making it possible for users to share with one 
another their recipes to build 3D printings. }
\end{itemize}

Obviously, the two factors are strictly connected in a closed loop: 
the more new users share their knowledge on building and using 3D printers, the more 3D printers' price 
will decrease of a magnitude's order. The phenomenon can be seen right now. 
Following the historical trend, nowadays the mean cost for 
a home 3D printer is around 3,500 USD while the first ones were 
of one or two order of magnitude more expensive. Originally 
built for the manufacturing industry, 3D printers are raising 
attention also in the biomedical field as a tool for producing 
biomedical membranes, pills and surgery devices remotely. In 
particular, the application of such technology is forcing new 
ways to approach the treatment of a patient both in hospitalization contexts in a laboratory 
as in home caring. In these notes, we will 
survey the currently most promising applications of 3D printers 
with an eye focused on the potentiality of these technologies 
for the telemedicine applications. The paper is organized as 
follows: Section II is dedicated to a brief description on how 
a 3D printer works. Section III will describe some of the most 
promising use of the 3D printers in the medical field. 
Section IV will briefly describe the 3D printers communities. The last section will describe some barriers that need to be addressed 
before a large acceptance of 3D printers in telemedicine could 
become reality. Some issues related to regulation will be addressed as well. 
\section{How 3D printer works}
It is not easy to collect in one single definition all of the 
technologies involved in the so called 3D print wave. Probably 
the largest part of the commercial 3D printers can be described 
as black-box home devices able to create solid objects made 
from powderer material. The suffix "printers'' in "3D printers'' relies on the fact 
that, from the user interface perspective, such devices work as 
common printers in a normal office. The device is connected 
(using USB) to a PC that codes a design into a series of processes 
that are sent to the device that outputs the object. The main types of 3D printing processes can be summarized as follows:

\begin{enumerate}
\item{Extrusion: uses plastic segment of a metal wire that is wound on a coil and unreeled to supply material to an extrusion nozzle.}
\item{Granular: uses selective fusion of materials in a granular bed. The granules are fused layer by layer until the object is built. }
\item{Laminated: Laminates objects using layers of thin plastic, paper or metal sheets}
\item{Light polymerized: 
Vat of liquid polymer is repeatedly exposed to light. The exposed liquid polymer hardens in small increments until the model has been built. 
The remaining liquid polymer is drained from the vat, leaving the solid model.
Another system sprays photopolymer materials in ultra-thin layers until the model is completed.
}
\end{enumerate}

 Nevertheless, considering the application filed, this is a very poor definition. Recently, 3D printing was applied to produce highly specialized electronic \cite{Ahn20032009}, microfluidic \cite{Therriault2003} and pneumatic devices \cite{ANIE:ANIE201006464}, \cite{Hasegawa2008390}. But these are still manufacturing related use of 3D printing technology, where the process involved is basically of physical type meaning that physical know processes are wired into a home usable device. Some major breakthroughs have been presented with the seminal paper by Symes et al. \cite{enlighten68744} that use a 3D printer for controlling chemical synthesis. 
They use the Rhino3D package and a low-cost (200 USD) Fab@home robocasting platform to create and control the synthesises and crystallization of two different polyoxometalates using a camera to control the reaction. The most common 3D printers are the ones that use extrusion to create plastic manufactured objects.

\section{Use of 3D printers in telemedicine:recent trends}
Probably the most interesting use of classical 3D printer in telemedicine application has been enlightened in a recent study \cite{doi:10.1056/NEJMc1206319} presented in the New England Journal of Medicine. This study is particularly interesting because it was the first out-of-the-lab use of a 3D printer in a surgical context. A newborn was diagnosed with Tracheobronchomalacia \cite{carden2005tracheomalacia} which is hard to treat and rapidly conduce to airway collapse and respiratory insufficiency. 
At the age of 20 weeks the baby's trachea was patched with a trachea splint, to allow normal flow ventilation. The splint was created from a biopolymer called polycaprolactone using a 3D printer. The device was created directly from a CT scan of the baby's trachea/bronchus, integrating an image-based computer model with laser-based 3D printing to produce the splint (see Fig 2).
\begin{figure}[htbp]
\centering
\includegraphics[width=7.5cm]{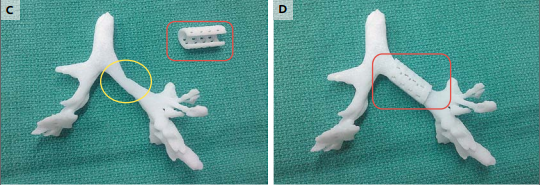}
\caption{Tracheal splint and the patched trachea \cite{doi:10.1056/NEJMc1206319} (p. 2044) }\label{figure:2}
\end{figure}
\newline
This first remote surgery splint creation using 3D printer can be seen as an astonishing potentially new use of 3D printers particularly in development countries where due to the lack of infrastructure for delivery medical prosthesis it is sometime more easy to have Global System for Mobile Communications (GSM) networks availability than fast medical device deliveries. For a strange paradox there are for example African countries where the mobile availability is greater than the available driveways \cite{Buys20091494}, and even countries where telemedicine applications have been delivered successfully, for example in malaria monitoring \cite{citeulike:12423394}.
\newline
\indent One of the most interesting cases has been described by Tam et al. \cite{JRCR889}. 
The surgery involved a 6 year-old girl with a large scapular osteochondroma complicating congenital diaphyseal aclasia. Osteochondroma is a type of benign tumor that consists of cartilage and bone. It is a benign cartilage-capped outgrowth, connected to bone by a stalk.
It is the most frequently observed neoplasm of the skeleton.
They generally occur at the end of the growth plates of long bones, often at joints. They most commonly form at the shoulder or the knee but have been known to occur in the long bones of the forearm (i.e. the radius and ulna). In this case, the girl had also a congenital diaphyseal aclasia that is a relatively rare abnormal condition that affects the skeletal system. Characterized by multiple exostoses or bony protrusions, it is inherited as a dominant trait. To help clinician visualize a 3D model of the tumor before going in vivo with the patient, a 3D model of the scapula was created by post-processing the Digital Imaging and COmmunications in Medicine (DICOM) image \cite{JRCR889}. Nowadays, DICOM files are a well established standard way of manipulating high resolution images of the human body as output of computed tomography or computed radiography \cite{citeulike:225180}. In this case, the 3D printer was used to create a 1:1 3D model of the girl's tumor  to help the clinician visualize it and test the procedure to adopt before entering into the operating room (see Fig 3.).
\begin{figure}[htbp]
\centering
\includegraphics[width=2.65cm]{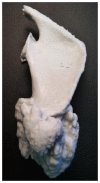}
\caption{3D model of scapula \cite{JRCR889} (p. 35)}\label{figure:3}
\end{figure}
\newline

This first step clearly leads to the other still promising use of this technology, that is, the 3D printing of human tissue for implantation purpose. A living organ, such as a liver or the heart itself, is too complex to reproduce as a single piece outside its connection to the other organs. However, one promising line for 3D printing is the production of human bones \cite{Fedorovich2011601}. Even if the human bones are a living structure, the fact that some bones replacements like hip replacement are becoming part of the standard surgery methodology for well known clinical protocols in ageing related pathologies \cite{hipreplacement} has driven research in the area of 3D printers.
\newline
\indent
In 2011, Anthony Atala \cite{TRI:TRI1182} took to the stage at the Technology, Entertainment, Design (TED) 
conference \cite{tedtalks} and showed the world a 3D printed kidney.
Atala's original 3D printed kidneys were made with a bio-ink that perfectly replicated kidney tissues, the problem was that these tissues were not vital (living). Without entering too much into details, the process involved stem cells that have the ability to transform themselves into other cells like nephrons, neurons and cardio muscles' cells. This pluri-potential cells are cultivated in a solution with a structure as support to allow them to aggregate in a structured way. As done in other contexts, the cells were forced to mutate to the desired ones and forced to aggregate in a structured way. The networking relations that exist in a human living kidney were lost. Without the ability to create living organs, 3D printed transplants would remain impossible, even if this step was a great breakthrough, for the potential implications of this technology in everyday life. Even more surprisingly, in 2013 Manoor et al. \cite{doi:10.1021/nl4007744} uses a 3D print to aggregate cells over the geometry of a human ear. 
The result is quite impressive (see Fig 4.). 
\begin{figure}[htbp]
\centering
\includegraphics[width=5.65cm]{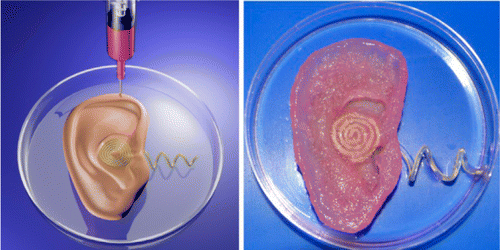}
\caption{3D printed bionic ear \cite{doi:10.1021/nl4007744} (p. 7)}\label{figure:4}
\end{figure}

So, even if creating from scratch a fully functional living organ to be used for 	
transplantation purpose is far away from being realized, this first step is quite astonishing. Outside the lab, a goal easier to reach would be the remote creation of prosthesis made by atossic polymer material. 
Last but not least, another promising approach to the use of 3D printer technology remotely involves the field of pharmacy production. With his seminal paper \cite{enlighten68744} professor Leroy Cronin demonstrates the possibility to create complex chemical reactions using a modified 3D printer. It was one of the first tries to initiate chemical reactions by printing (i.e. producing) the reagents directly into a 3D reactionwave matrix. Using this approach it is possible to control, with a software, the reactionware design, construction and operation. Another interesting fact is that the whole proof of concept was created using a low-cost 3D printer (approx. 2,000 USD) and open-source design software (see Fig 5). 
\begin{figure}[htbp]
\centering
\includegraphics[width=5.65cm]{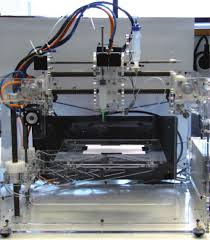}
\caption{Open-source design 3D printer}\label{figure:4}
\end{figure}
\section{Online 3D printers community}
In the last section, we have seen some of the recent trends in using advanced 3D printing techniques. Obviously, some of them are too complicated and they build materials so difficult to manage that it is not possible to think of them outside a controlled lab environment. Despite this, standardized 3D printing methods using granular plastic material and estrusors to create plastic manufactured object are right now a reality. This technology is so well established that entire web sites like Thinkverse \cite{thingiverse} have been created to share the design for the objects to be printed. So, one user can download the design schema for a specific model of 3D printer to build a plastic made object. The community itself in this case follows up the line of the open-source so that every member of the community is supported to share his/her own design with the other community members. Every member can upload the instruction and design to be downloaded by other users who want to build the object. The revenue for this process is based on the fact that every new recipe and built object is immediately shared between site surfers without any need of registration. This means that every newcomer can build objects without the need of a registration; this step is required only if he/she has new/modified recipes to share.  
Such communities themselves are now starting to be the object of research \cite{Mota:2011:RPF:2069618.2069665}. The interaction on these communities is driven in some ways by a hacking spirit.The next step in the area of 3D printer communities will be to go from the pure technical issues to applications in the medical area. These communities will attract researchers, designers and clinicians who want to exchange ideas and experiences from their everyday situations. It becomes clear from the development of 3D printing in the health care that it is vital to keep a viable conversation about hygiene factors, including critical thinking of existing processes, how to handle materials, and so on. Online communities-of-practice for collaborative learning could be about how to produce 3D objects in a safe and hygienic way, according to standardized routines. 
\section{Future possibilities and the regulation}
As usual, when technology is running so fast in a way that has been defined as garage-science, the legislature is in trouble chasing the different fast changes.
If one may dream, for example, of a 3D printer for building prothesis for human livings, according to most strict interpretation of the international law, we need not only to guarantee the safeness of the whole production process, but also to ensure sanitary standards that we normally find in hospitals and biomedical manufacturing environments. 
In the case presented by Zopf et al. \cite{doi:10.1056/NEJMc1206319}, we notice also that, before doing the surgery, the clinicians need to have an emergency clearance from the Food and Drug Administration (FDA), being that the polycaprolactone biopolymer does not have consensus to be used by the FDA. The production of medical device inside the US is strictly regulated by the FDA. 
So, before entering the marketing stage, one medical device needs to be certified by the FDA. However, for a medical device that does not appear in the FDA medical device database \cite{FDAMEDDEV} the use is possible in particular situations \cite{FDA}. In one case, even though the procedure and the material was not intended to be used on human being, the surgery could take place because it was considered as compassionate cure.
Outside this context that is somehow a life risk situation, in normal medical device manufacturing there is both in EU and US a strict regulation that assures the safeness of the product itself both for the patient and for the clinician. In the particular cited case the problem was somehow bypassed by the fact that the nursing process took place in the US territory under the auspices of the same agreed regulation. This leads to a potential interesting law problem, as usual when dealing with remote assistance. If we imagine a remote extrusion 3D printer that uses a polymer to build a splint in a region outside the US, what should be the best way to assure the patient the safeness of the process and to reduce at minimum the risk of rejection by the patient? And upon this, in case something goes wrong, who is the actor being responsible and for what is he/she responsible? As in most of the latest technology breakthroughs, for the moment, the technology wave innovation is leading us to new and unseen possibilities, and not only for the western countries. Once the tide will lower a little, a regulation should be introduced to manage the issues arising from the adoption of this new technology. The time is approaching, because in February 2014, key patents that currently prevent competition in the market for the most advanced and functional 3D printers will expire. When this will happen, when the key patents on 3D printing via laser sintering will expire, we will most likely see a huge drop in the price of these devices. This just happened, when the key patents expired on a more primitive form of 3D printing, known as fused deposition modelling (FDM). The result was an explosion of open-source FDM printers that eventually led to iconic home and hobbyist 3D printer manufacturers. When the medical use of 3D printers becomes widely spread, it is time to initiate conversations about the practitioner's work with 3D printers. Also, systematic evaluations of the use of 3D printers will be beneficial to the area.

\section*{Acknowledgment}

The first author would like to thank Giuseppe Dalla Motta for his precious discussions on the chemical tasks involved in the industrial 3D printing process.



\bibliographystyle{IEEEtran}
\bibliography{etelemed2014}

\end{document}